\documentclass{pasa}%

\usepackage{graphicx}
\newcommand{\pasa}{Publications of the Astronomical Society of Australia}

\newcommand{\aap}{Astronom. and Astrophys.}

\newcommand{\aaps}{Astronom. and Astrophys. Suppl. Ser.}

\newcommand{\aj}{Astronom. J.}
\newcommand{\apj}{Astrophys. J.}

\newcommand{\apss}{Astrophys. and Space Sci.}

\newcommand{\mnras}{Monthly Notices Roy. Astronom. Soc.}

\def\farcs{\hbox{$.\!\!^{\prime\prime}$}}
\def\fm{\hbox{$.\!\!^m$}}
\def\fs{\hbox{$.\!\!^s$}}
\def\degr{\hbox{$^\circ$}}

\title[Complex Analysis of the Stellar Binary HD25811]{Complex Analysis of the Stellar Binary HD25811; A Subgiant System}
\author[Al-Wardat et al.]{ Mashhoor A. Al-Wardat$^1$, Hatem S. Widyan$^2$ \and Ahmed Al-thyabat$^{2,}$\thanks{mwardat@ahu.edu.jo}\\
\affil{$^1$Department of Physics, Al-Hussein Bin Talal University, P.O.Box 20, 71111, Ma'an, Jordan.}%
\affil{$^2$Department of physics, Al al-Bayt University, Mafraq, Jordan}}%
\jid{PASA}
\doi{10.1017/pas.\the\year.xxx}
\jyear{\the\year}

\usepackage[authoryear]{natbib}
\bibpunct{(}{)}{;}{a}{}{,}
\begin{document}%
\begin{abstract}
The visually close binary system HD25811 is analyzed to estimate its physical and geometrical parameters in addition to its spectral type and luminosity class.
The  method depends on obtaining the best fit between the entire observational spectral energy distribution (SED) of
the system and synthetic SEDs created by atmospheric modeling of the individual components, consistent with the system's modified orbital elements.
The parameters of the individual components of the system are derived as: $T_{\rm eff}^{\rm a}
=6850\pm50$\,K, $T_{\rm eff}^{\rm b} =7000\pm50$\,K, log $g_{\rm a}=4.04\pm0.10$,
 log $g_{\rm b}=4.15\pm0.10$, $R_{\rm a}=1.96\pm0.20$\,R$_{\odot}$,  $R_{\rm b}=1.69\pm0.20$\,R$_{\odot}$, $M_{v}^{\rm a}=1\fm97\pm0.20$, $M_{v}^{\rm b}=2\fm19\pm0.20$, $L_a= 7.59\pm0.70 L_\odot, L_b= 6.16\pm0.70 L_\odot$ with dynamical parallax $\pi(\textrm{mas})=5.095\pm 0.095$. The analysis shows that the system consists of a $1.55M_{\odot}$ F2 primary star and a less evolved $1.50M_{\odot}$ F1 secondary subgiant star with ages around 2 Gy formed by fragmentation. Synthetic magnitudes of both components were calculated under Johnson-Cousins, Str\"{o}mgren, and Tycho photometrical systems.
\end{abstract}

\maketitle%

\section{INTRODUCTION }
Analysis of stellar binary systems is the most reliable and accurate way of estimating stellar physical and geometrical parameters, especially stellar masses. It emphasizes the role of binary stars in examining some physical and stellar evolutionary theories.
Even high resolution observational techniques, like speckle
interferometry and adaptive optics, are not sufficient to
determine the physical parameters of the  individual components of visually close (spatially unresolved on the sky) binary systems (VCBS), especially in the case of this study (HD25811) which has no parallax measurement.

Combining  spectrophotometry with atmospheric modeling gives a new method for the accurate determination of the physical and geometrical parameters of both components of VCBS, in addition to the estimation of their spectral types, luminosity classes and ages.  This method, first advised by \cite{2002BSAO...53...51A, 2007AN....328...63A},  was applied to several VCBS such as ADS11061, Cou1289, Cou1291, Hip11352, Hip11253, Hip70973 and Hip72479  \citep{2002BSAO...53...51A, 2007AN....328...63A, 2009AN....330..385A, 2012PASA...29..523A, 2009AstBu..64..365A}.

In this paper, we are developing the method by combining the dynamical analysis of the relative orbit of the binary, a step which will strengthen the method, reduce the error bars and make it applicable to evolved binaries by accurate determination of the individual masses.

The system HD 25811 (SAO 93759 = CHA13 = BAG~4) was first
resolved as a binary using the lunar-occultation observational
technique in the early eighties \citep{1984AJ.....89.1371S, 1985AJ.....90.2360E}. Subsequently it was included in the  speckle
interferometric program of the Russian 6-meter telescope by \cite{1987PAZh...13..508B}, and was designated as BAG~4. Since then it is
routinely observed by speckle interferometric techniques all over
the world to achieve as much data as possible in order to
calculate its best orbit \citep{1987AJ.....93..688M, 1989AISAO..28..107B, 1989AJ.....97..510M, 1990AJ.....99..965M, 1992AJ....104..810H, 1992AJ....103.1399B, 1994A&AS..105..503B, 2001AstL...27...95B, 2007AstBu..62..339B}(see Table \ref{pointsBAG4}).  But it was not listed in the Hipparcos program, which is why it has no trigonometric parallax measurement.

The VCBS HD25811 (BAG~4) fulfills the requirements of the aforementioned method, where it has  precise magnitude difference measurements and  observational spectral energy distribution (SED) covers the optical range. It has also several relative positional measurements listed in the Fourth Catalog of Interferometric Measurements of Binary Stars (\textit{http://ad.usno.navy. mil/wds/int4.html}), which they used to build its preliminary orbit and a new positional measurement that can be used to modify it.

 In addition to that, HD26811 represents  a very good example for studying the formation
and evolution of stellar binary systems, since, as we shall prove, it consists of two subgiant stars.

Table~\ref{table1} contains the basic data of the system from SIMBAD and other databases, and Table~\ref{table2} contains data from  Tycho-2 Catalogue \citep{1997yCat.1239....0E}.

\begin{table}
\begin{center}
\caption{Data from SIMBAD and NASA/IPAC} \label{table1}
\begin{tabular}{lcc}\hline \hline
  & HD25811  & source
   \\ \hline
$\alpha_{2000}$ & $04^h 06^m 16\fs416$ &1\\
$\delta_{2000}$&$+19\degr52' 28.''57$&1\\
 Tyc &  1258-22-1 &1\\
 HD &  25811   &1\\
 Sp. Typ. & F0 &1\\
 E(B-V) &0.278&2\\
 $A_v$&$0\fm870$&2\\
\hline \hline
\end{tabular}
\\
$^1${SIMBAD}, $^2${NASA/IPAC:http://irsa.ipac.caltech.edu},
\end{center}
\end{table}
\begin{table}[!ht]
\begin{center}
\caption{Data from Tycho-2 Catalogue \citep{2000A&A...355L..27H}}
\label{table2}
\begin{tabular}{lcc}\hline \hline
  & HD25811  
  \\ \hline
 $V_J(Hip)$ & $8\fm66$
 \\
 $B_T$ & $9\fm11\pm0.020$ 
 \\
 $V_T$ & $8\fm71\pm0.017$ 
 \\
 $(B-V)_J(Tyc)$ & $0\fm389\pm0.028$ 
 \\
\hline \hline
\end{tabular}
\end{center}
\end{table}

\section{ORBITAL ELEMENTS}
\label{orbital_elements}
The  orbit of the system  was calculated firstly by
\cite{2001AstL...27...95B} using the first 15 relative positional measurements in  Table ~\ref{pointsBAG4}, they noted that they will have to wait another $\sim 10$ years to obtain a reliable solution. Then it was modified by \cite{2003PhDT.......174G} using the  measurements up to 2002.797. A new slight modification of the orbit is introduced here using all relative positional measurements listed in Table ~\ref{pointsBAG4}, which cover around $210^\circ$ of the complete orbit. 

\begin{table*}
\begin{center}
\caption{Estimated orbital elements  of the system with those of the old orbits by  \citep{2001AstL...27...95B} and \citep{2003PhDT.......174G}.}
 \label{orbital_elements}
\begin{tabular}{lccc}
\hline \hline
  Parameters & This work & \citep{2003PhDT.......174G} &\citep{2001AstL...27...95B}\\

 \hline
 Period $P$                 & $32.99^y\pm 2.37^y$     & $31.00^{y} \pm 1.33^{y}$                   & $30.68^{y} $\\
 Periastron epoch  $T_0$   &   $ 1993.26 \pm 18.76$ & $1981 \pm 12$                       & $1990.96 $      \\
 Eccentricity $e$           & $    0.047 \pm 0.040$& $0.007 \pm 0.032$                        & $0.045 $        \\
  Semi-major axis $a$       & $    0\farcs076 \pm  0\farcs002$& $0\farcs076 \pm 0\farcs001$           & $0\farcs079  $ \\
  Inclination $i$           & $123^\circ \pm 3^\circ$& $ 124^\circ \pm 2^\circ$               & $ 128^\circ $  \\
 Argument of periastron $\omega$    & $68^\circ \pm 216^\circ$& $291^\circ \pm 140^\circ$        & $32^\circ $   \\
 Position angle of nodes $\Omega$   & $59^\circ \pm 3^\circ$& $61^\circ \pm 1^\circ$          & $50^\circ $   \\
   \hline \hline
\end{tabular}\\
\end{center}
\end{table*}

The estimated orbital parameters  of the system along with those of the old
orbits are listed in Table ~\ref{orbital_elements}.

\begin{figure}[!h]
\resizebox{\hsize}{!} {\includegraphics[]{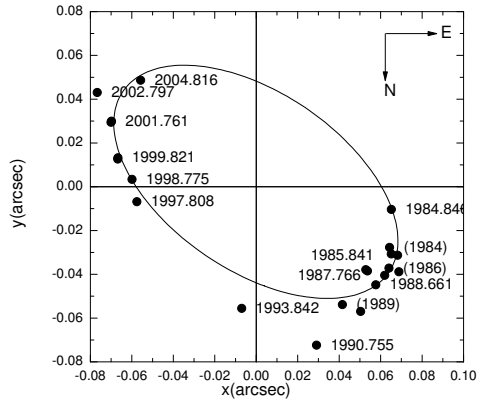}}
 \caption{Relative visual orbit of the system with the epoch of the positional measurements; the origin represents the position of the primary component.}
 \label{orbit}
\end{figure}

\begin{figure}[!h]
\resizebox{\hsize}{!} {\includegraphics[]{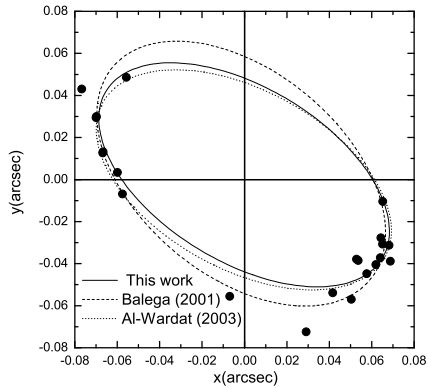}}
 \caption{Comparison between the modified relative visual orbit  of the system in this work (solid line) and those of \cite{2001AstL...27...95B} (dashed line) and  \cite{2003PhDT.......174G} (doted line).}
 \label{orbit_all}
\end{figure}

Fig. ~\ref{orbit} shows the relative visual orbit of the system with the epoch of the positional measurements, and Fig. ~\ref{orbit_all} shows the new orbit against the old orbits of \cite{2001AstL...27...95B} and \cite{2003PhDT.......174G}.

Unfortunately there is no any positional measurement for the system during the last 9 years, which would have covered the whole orbital period. This could give a more reliable orbit and hence more precise geometrical parameters.


\begin{table}
\begin{center}
\caption{Relative positional measurements using different methods which are used to build
the orbit of the system. These points are taken from the Fourth Catalog of
Interferometric Measurements of Binary Stars and a point from  \cite{2003PhDT.......174G}. }
\label{pointsBAG4}
\begin{tabular}{lcccc}\hline \hline

    Date    &  $ \theta$(deg)&   $\rho$(deg)  &   Ref. & Meth.   \\
\hline
  1984.1127 &  66.7    &       0.0701  &    Smk1984 &  Occ\\
  1984.1127 & 254.4     &      0.0653  &   Evn1985  & Occ\\
  1984.1127 & 254.4     &      0.0663  &    Evn1985 & Occ\\
  1984.8460 &  81.0     &     0.066   &     Bag1987&  Spe\\
  1985.8406 &  65.4     &      0.075    &    McA1987b& Sch\\
  1986.6573 &   64.8     &      0.072   &   Bag1989a &Spe\\
  1986.8862 &   59.9     &      0.074   &   McA1989  &Sch\\
  1986.8890 &   60.6     &      0.079   &    McA1989 & Sch\\
  1987.7655 &   56.9     &      0.074   &   McA1989  &Sch\\
  1988.6609 &   52.2     &      0.073   &     McA1990 & Sch\\
  1989.7067 &   41.5     &      0.076   &     Hrt1992b& Sch\\
  1989.8077 &   37.7    &      0.068    &    Bag1994  &Spe\\
  1990.7551 &   21.9     &      0.078   &    Hrt1992b &Sch\\
  1993.8419 & 353.0    &     0.056     &     Bag1994  &Spe\\
  1997.808  & 276.9    &     0.0582    &     Bag2001  &Spe\\
  1998.7747 & 266.9    &     0.060     &     Bag2002  &Spe\\
  1999.8185 &  259.4   &     0.0681    &     Bag2004  &Spe\\
  1999.8213 &  258.9   &     0.0683   &      Bag2004  &Spe\\
  2001.7614 &  247.3   &     0.076    &      Bag2006b &Spe\\
  2001.7614 &  246.9   &     0.076    &      Bag2006b &Spe\\
  2002.797  &   240.78 &   0.0880     &      Wrd2003& Spe  \\
  2004.8158 &  229.0   &     0.074    &      Bag2007b &Spe\\

\hline \hline
\end{tabular}
\end{center}
\tabnote{Occ: Occultaion}.
\tabnote{Spe: Speckle Interferometry}.
\tabnote{Sch: CHARA speckle}.
\tabnote{Wrd2003: \cite{2003PhDT.......174G}}.
\tabnote{References are abbreviated as in the Fourth Catalog of
Interferometric Measurements of Binary Stars}.
\end{table}

\section{ATMOSPHERIC MODELING}

To calculate the individual components'  preliminary input parameters ($T_{\rm eff}$ and $\log\,g$), we followed the procedures explained in \cite{2012PASA...29..523A} using the following relations \citep[e.g.,][]{1992adps.book.....L, 2005oasp.book.....G}:
\begin{eqnarray}
\label{eq2}
m_v=-2.5\log(f_1+f_2)\\
M_v=m_v+5-5\log(d)-A,\\
\label{eq4}
\log(R/R_\odot)= 0.5 \log(L/L_\odot)-2\log(T/T_\odot),\\
\label{eq5}
\log g = \log(M/M_\odot)- 2\log(R/R_\odot) + 4.43,
\end{eqnarray}
where  we used the bolometric corrections of
\cite{1992adps.book.....L} \& \cite{2005oasp.book.....G},  $T_\odot=5777 \rm{K}$ and extinction ($A_v$) given in Table~\ref{table1} by NASA/ IPAC.

 So, using the entire visual magnitude of the system $m_v=8\fm66$ from Table~\ref{table2} and $\triangle
m=0\fm23$ from speckle interferometric results (Table~\ref{deltam2}) as the average of $\triangle m$
measurement under the filter  $545nm/30$ (the closest  filter to the Johnson V filter), we got $m_{v\rm a}=9\fm304$ and $m_{v\rm b}=9\fm534$.


The dynamical parallax estimated by \cite{2003PhDT.......174G}($\pi=5.24 \pm 0.6, d=191$\,pc) is used as a preliminary value to calculate the individual absolute magnitudes according to the following equation:
\begin{eqnarray}
\label{eq3}
M_v=m_v+5-5\log(d)-A.
\end{eqnarray}
This leads to the following preliminary input parameters:
\begin{eqnarray}
\label{p1}
M_{v\rm a}=2\fm03, \,\,\,\,  M_{v\rm b}=2\fm26 \\
\label{p2}
T_{\rm eff}^{\rm a}=8150\,{\rm K}, \,\, \,  T_{\rm eff}^{\rm b}=7950\,{\rm K},\\
\label{p3}
\log\,g_{\rm a}=4.2, \,\,\, \log\,g_{\rm b}=4.2,
\end{eqnarray}
\noindent and Spectral Types  A6, A7.
Which in their turn used to build the preliminary line blanketed model atmospheres using ATLAS 9 \citep{1994KurCD..19.....K}.



\begin{table}
\begin{center}
\caption{Magnitude difference between the components of the
system, along with the filter used to obtain the
observations. } \label{deltam2}
\begin{tabular}{lcc}
\hline \hline
   $\triangle m $& filter ($\lambda/\Delta\lambda$)& ref.  \\
\hline
 $0\fm00\pm0.40$ & $ 545nm/30 $& 1 \\
  $0\fm18\pm0.12$ &$610nm/20 $  &2\\
  $0\fm23\pm0.27$ & $545nm/30 $  &2\\
  $0\fm22\pm0.20$ &$850nm/75 $  &3\\
  $0\fm21\pm0.17$ & $545nm/30 $  &3\\
  $0\fm24\pm0.27$ & $545nm/30 $  &4\\
 \hline \hline
\end{tabular}\\
\end{center}
\tabnote{$^1$\cite{2002A&A...385...87B}},
\tabnote{$^2$\cite{2004A&A...422..627B}},
\tabnote{$^3$\cite{2006BSAO...59...20B}},
\tabnote{$^4$\cite{2007AstBu..62..339B}}.
\end{table}

\subsection{Synthetic spectra}
\begin{figure}[!ht]
\resizebox{\hsize}{!} {\includegraphics[]{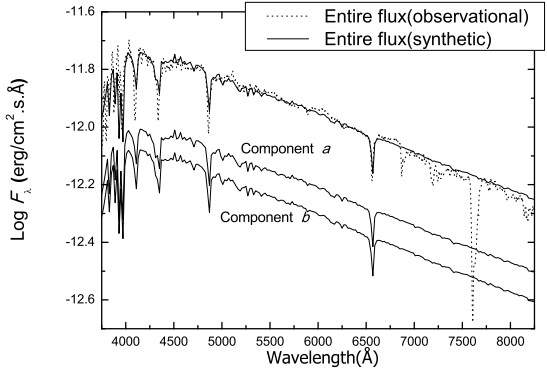}}
 \caption{Dotted line: observational SED in the continuous spectrum of the
 system. Solid lines: the  entire computed SED of the two components,
  the computed flux of the primary component with $T_{\rm eff}=6850\pm50$\,K,
 log $g=4.04\pm0.10, R=1.96\pm0.20R_\odot$, and the computed flux of the secondary
component with
 $T_{\rm eff} =7000\pm50$\,K, log $g=4.15\pm0.10, R=1.69\pm0.20 R_\odot $, and \,
$ d=196.27$\,pc.}
 \label{Modfit1291}
\end{figure}

The observational  SED of the entire system was taken from \cite{2002BSAO...53...58A} (Fig.~\ref{Modfit1291}).
It shows some strong lines and depressions, especially in
the red part of the spectrum (around
$\lambda$6867\AA, $\lambda$7200\AA, and $\lambda$7605\AA). These are
$\rm H_2O$ and $\rm O_2$ telluric lines and depressions.
Synthetic  Jonson-Cousins, Str\"{o}mgren and Tycho magnitudes and colour indices of this observational  SED  are listed in Table~\ref{obsmag}.

Synthetic SEDs of the individual components are built first using solar metalicity  model atmospheres as the output of ATLAS 9 of the preliminary parameters (Equs. ~\ref{p1},~\ref{p2}and ~\ref{p3}), $R_{\rm a}=1.55\,{\rm R}_\odot, R_{\rm b}=1.45\,{\rm R}_{\odot}$  and $d=191$pc (from \cite{2003PhDT.......174G}).  The total energy flux from a binary star is created from the net
luminosities of the components $a$ and $b$ located at a distance $d$ from the Earth as follows:
\begin{equation}
 F_{\lambda} \cdot d^2 = H_{\lambda}^{\rm a} \cdot R_{\rm a}^2 +
 H_{\lambda}^{\rm b} \cdot R_{\rm b}^2\,,
\end{equation}
from which
\begin{equation}
\label{eqFlux}
F_{\lambda}  = (R_{\rm a}^2/d)^2(H_{\lambda}^{\rm a} + H_{\lambda}^{\rm b}
\cdot(R_{\rm b}/R_{\rm a})^2)\,,
\end{equation}
where $H_{\lambda}^{\rm a}$ and  $H_{\lambda}^{\rm b}$ are the fluxes from a
unit surface of the corresponding component. $F_{\lambda}$ here
represents the entire SED of the system; which for a given  $H_{\lambda}^{\rm a,b}$ depends on the parallax of the system and the radii of the components.

\begin{table}
\begin{center}
\caption{Entire synthetic Johnson-Cousins, Str\"{o}mgren and Tycho
magnitudes and  colour indices of the system HD25811
\citep{2008AstBu..63..361A}.} \label{obsmag}
\begin{tabular}{lccccc}
\hline  \hline
\noalign{\smallskip}
  System & Filter& HD25811 
   \\
 \hline
Johnson-Cousins &$B_J$ & $9\fm05$ 
      \\
      &$V_J$ & $8\fm64 $ 
      \\
       &$(B-V)_J$ & $0\fm41$
      \\
    \hline

Str\"{o}mgren &$v$ & $9\fm31$ 
          \\
          &$b$& $8\fm91$ 
          \\
             &$y$ & $8\fm61$ 
          \\
         &$v-b$& $0\fm40$ 
        \\
        & $b-y$ & $0\fm30$  
        \\

   \hline
    Tycho       &$B_T$ & $9\fm14 $ 
              \\
              &$V_T$ & $8\fm70$
                 \\
              &$(B-V)_T$ & $0\fm44$ 
                \\
         \hline \hline
\end{tabular}
\end{center}
\end{table}

The resultant entire synthetic SED did not fit the observational SED. Therefore, many attempts were made and hundreds of synthetic SED's were built, using different sets of parameters, and compared with the observational SED to achieve the best fit. The attempts included changing the effective temperatures, gravity acceleration, parallax and radii. It is worthwhile  mentioning  that the step size of the change was less than or equal to the estimated error in each parameter. In the case of the effective temperatures and radii, the changes exceed the errors of the first estimations because we are dealing with sub-giants instead of main sequence, or in other words, that is what we discovered.

Kepler's equation is used to match between the physical and geometrical parameters, and to connect the dynamical analysis with the atmospheric modeling:
  \begin{eqnarray}
\label{eq9}
\ (\frac{M_a + M_b}{M_\odot})(\pi^3)=\frac{a^3}{p^2},
\end{eqnarray}
\noindent where $(\frac{M_a + M_b}{M_\odot})$ is the mass sum of the two components in terms of solar mass, $\pi$ is the parallax of the system in arc seconds, $a$ is the semi-major axes in arc seconds and $p$ is the period in years.

Fixing the right part of  equation ~\ref{eq9} according to the orbital elements of the system calculated in section \ref{orbital_elements}, and changing the values of the left part in an iterated way,  ensures agreement between the masses of the components, their positions on the evolutionary tracks and the best fit between the synthetic and observational SED's.

For stars with formerly estimated temperatures, masses and gravity accelerations, this agreement is possible only for radii bigger than what would be if they were main sequence stars (the preliminary input radii), which means that both components are evolved subgiant stars.

Within the criteria of the best fit, which are the maximum values of the absolute fluxes, the inclination of the continuum of the spectra, and the profiles of
the absorption lines, the best fit  (Fig.~\ref{Modfit1291}) is found using the following
set of parameters:
\begin{equation}
T_{\rm eff}^{\rm a} =6850\pm50\,{\rm K}, \
T_{\rm eff}^{\rm b} =7000\pm50\,{\rm K},
\end{equation}
\begin{equation}
\log g_{\rm a}=4.04\pm0.10, \ \log g_{\rm b}=4.15\pm0.10,
\end{equation}
\begin{equation}
R_{\rm a}=1.96\pm0.20\,{\rm R}_{\odot}, R_{\rm b}=1.69\pm0.20\,{\rm R}_{\odot},
 \label{Rab}
\end{equation}
\begin{equation}
(M_a + M_b)/M_\odot = 3.05\pm 0.10,
\end{equation}
 and
$5.095\pm 0.095, d=196.27$\,pc.

The role of Equation \ref{eq9} rises here to assure the compatibility between the parallax of the system and its mass sum being given the period and the semi-major axes.

The complete set of the physical and geometrical parameters are listed in Table ~\ref{tablef1}.

The estimated parameters are highly dependent
on the precision of observations. Thus, within the error values of the measured quantities, they represent adequately  enough  the
parameters of the systems' components.

Depending on the Tables of \cite{1981Ap&SS..80..353S} and \cite{2005oasp.book.....G},  the spectral types for the components of the system are estimated as F2 IV for the primary component  and F1 IV for the secondary
component.

\section{SYNTHETIC PHOTOMETRY}
In addition to its importance in calculating the entire and individual synthetic magnitudes of the system, synthetic photometry  is used here as an evaluation technique to examine the best fit between the synthetic and observational spectra. This is performed by comparing the observed  magnitudes of the entire system from different sources  with the entire synthetic ones using  the following relation  \cite{2006AJ....131.1184M}
\cite{2007ASPC..364..227M}:
\begin{equation}
m_p[F_{\lambda,s}(\lambda)] = -2.5 \log \frac{\int P_{p}(\lambda)F_{\lambda,s}(\lambda)\lambda{\rm d}\lambda}{\int P_{p}(\lambda)F_{\lambda,r}(\lambda)\lambda{\rm d}\lambda}+ {\rm ZP}_p\,,
\end{equation}
 where $m_p$ is the synthetic magnitude of the passband $p$, $P_p(\lambda)$ is the dimensionless sensitivity function of the passband $p$, $F_{\lambda,s}(\lambda)$ is the synthetic SED of the object and $F_{\lambda,r}(\lambda)$ is the SED of the reference star (Vega).  Zero points (ZP$_p$) from  \cite{2007ASPC..364..227M} (and references therein) were adopted.

 Table~\ref{synth2} shows the magnitudes and colour  indices of the entire system and individual components in $UBVR$ Johnson-Cousins, $uvby$ Str\"{o}mgren and $BV$ Tycho.

\begin{table}[!ht]
\small
\caption{Synthetic magnitudes and colour indices  of the  system.}
\label{synth2}
\begin{tabular}{lcccc}
\hline\hline
Photomet. & Filter & entire & comp.  & comp.   \\
    System  &     &    $\pm0.03$   &    a     &    b      \\
\hline
Johnson-        & $U$ & $9.04$ & $9.70$ & $9.89$ \\
 \, Cousins     & $B$ & 9.04          &  9.70 &  9.89  \\
               & $V$ & 8.66          &  9.31 &  9.53 \\
               & $R$ & 8.44          &  9.08 & 9.32  \\
               &$U-B$& 0.00           & 0.00 & 0.00 \\
               &$B-V$&0.38              &  0.40 & 0.36 \\
               &$V-R$& 0.22             &  0.23 & 0.21 \\
  \hline
Str\"{o}mgren        & $u$ & 10.25 & 10.91 &  11.11  \\
                    & $v$ & 9.30  & 9.96  &  10.14 \\
                    & $b$ & 8.88 & 9.54 &  9.74 \\
                    &  $y$& 8.64 & 9.28 &  9.51  \\
                    &$u-v$& 0.96 & 0.95&  0.97 \\
                    &$v-b$& 0.42& 0.43 & 0.40 \\
                    &$b-y$& 0.25& 0.25& 0.24 \\
  \hline
  Tycho       &$B_T$  & 9.13 & 9.79 & 9.98   \\
              &$V_T$  & 8.71 & 9.36 & 9.58  \\
              &$B_T-V_T$& 0.42& 0.44& 0.40\\
\hline \hline
\end{tabular}
\end{table}

\section{RESULTS AND DISCUSSION}

 Comparing the  synthetic magnitudes and colours  with the observational ones (Table~\ref{tablecopmarison})  shows a high  consistency between them. This gives a good indication about the reliability of the  estimated parameters of the individual components of the system which are listed in Table~\ref{tablef1}.

\begin{figure}[!ht]
\resizebox{\hsize}{!} {\includegraphics[]{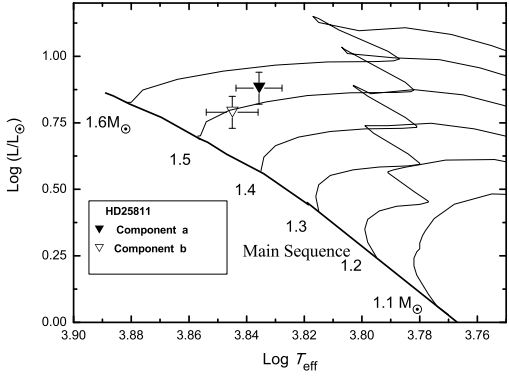}}
 \caption{Components of the system on the evolutionary tracks of Girardi et al.  \cite{2000A&AS..141..371G}. }
 \label{evol2}
\end{figure}

 Fig.~\ref{evol2} shows the positions of the system's components on the
 evolutionary tracks of   \cite{2000A&AS..141..371G}. It shows  that the primary more massive component is more evolved than the secondary, which is why the estimated value of the luminosity of the primary is  greater than the expected value for a MS star of the same mass.

 The estimated parameters of the system (Table~\ref{tablef1}) show that the  components of the system are  similar to the  star $\beta$ Hydri (HIP 2021), which is a G2IV evolved subgiant with an age of about 6.5 - 7.0 Gyr  \citep{1998A&A...330.1077D, 2003A&A...399..243F}, $\overline{\rho}(\overline{\rho}_\odot)=0.1803\pm0.0011$ (\cite{2007ApJ...663.1315B} using high precision asteroseismology)  and $T_{\rm eff}(K)=5872\pm44$, $R(\textrm{R}_{\odot}=1.814\pm0.017$), $\log g=3.952\pm0.005$ $L (\textrm{L}_\odot)=3.51\pm0.09 $ and  mass $M (\textrm{M}_{\odot})=1.07\pm0.03$ (\cite{2007MNRAS.380L..80N} using interferometry).

  This, in addition to their luminosity-temperature relation which reflects their positions on the HR diagram and the evolutionary tracks (see Fig. \ref{evol2}), their absolute magnitudes  which are brighter than those of MS stars of the same temperatures, and their densities, leads us to conclude that both components are subgiant stars.

 The age of the system was established from the evolutionary tracks as almost 2 Gy.  The similarity between the two components
  leads us to adopt the fragmentation  process for the formation of the
system, since it is the most likely mechanism in this case (for
more discussion, see \cite{2001IAUS..200...23B}, \cite{1975MNRAS.172P..15F}, and \cite{1987gady.book.....B}).


\begin{table}[!ht]
\small
\begin{center}
\caption{Comparison between the observational and synthetic  magnitudes, colours and magnitude differences of the system.}
\label{tablecopmarison}
\begin{tabular}{lcc}
\hline \hline
              & $\textrm{Obs}.^*$    & Synth. (this work)  \\
\hline
\noalign{\smallskip}

  $V_J$     & $8\fm66$          & $8\fm66\pm0.03$  \\
  $B_T$     & $9\fm11\pm0.020$ & $9\fm13\pm0.03$ \\
  $V_T$     & $8\fm71\pm0.017$ & $8\fm71\pm0.03$ \\
  $(B-V)_J$ & $0\fm389\pm0.028$ & $0\fm38\pm0.04$ \\
  $\triangle m $& $0\fm23^\dag$ & $0\fm22\pm0.04$\\

\hline \hline
\end{tabular}\\
\end{center}
$^*${See Table~\ref{table2}}.\\
$^\dag${Average value of the observational measurements using the filter  $545nm/30$ (Table~\ref{deltam2}) }.
\end{table}

\begin{table}[!ht]
\small
\begin{center}
\caption{Physical and geometrical parameters of the system HD25811 components.}
\label{tablef1}
\begin{tabular}{lcc}
\hline \hline
Component & a &  b  \\
\hline
\noalign{\smallskip}
$T_{\rm eff}$\,(K) & $6850\pm50$ & $7000\pm50$ \\
Radius (R$_{\odot}$) & $1.96\pm0.20$ & $1.69\pm0.20$ \\
$\log g$ & $4.04\pm0.10$ & $4.15\pm0.10$ \\
$L (L_\odot)$ & $7.59\pm0.70 $  & $6.16\pm0.70$\\
$M_{V}$ & $1\fm97\pm0.20$ & $2\fm19\pm0.20$\\
Mass, ($M_{\odot})$& $1.55\pm0.16$ & $1.5 \pm0.15$  \\
$\overline{\rho}(\overline{\rho}_\odot)$& $0.21\pm0.05$& $0.31\pm0.05$\\
Sp. Type$^*$ & F2 IV & F1 IV \\
\hline
\multicolumn{1}{l}{Parallax (mas) }    & \multicolumn{2}{|c}{$5.095\pm 0.095$}\\
\multicolumn{1}{l}{$(M_a + M_b)/M_\odot$ }    & \multicolumn{2}{|c}{$3.05\pm 0.10 $}\\
\multicolumn{1}{l}{Age (Gy)}    & \multicolumn{2}{|c}{$2.0\pm 0.4 $}\\

\hline \hline
\end{tabular}
\end{center}
\end{table}

\section{CONCLUSIONS}

The VCBS HD25811 is analyzed using a modified version of the complex method which was first advised by \cite{2002BSAO...53...51A, 2007AN....328...63A}.

The modification in the method includes the dynamical analysis of the relative orbit of the system (which is very helpful in the determination of the masses  of the individual components of the system) and consequently their spectral types, by combining them with the atmospheric modeling input parameters.

 The physical and  geometrical parameters of the system's  components are estimated  depending on the best fit between the observational and
  synthetic SEDs, built using the atmospheric modeling of the individual components and the system's orbital elements.

Both components of the system  are concluded to be F2 IV for the primary and F1 IV  for the  secondary.

    The entire and individual $UBVR$ Johnson-Cousins, $uvby$ Str\"{o}mgren and $BV$ Tycho synthetic magnitudes and colours of the system are calculated.

    Finally, fragmentation  was proposed as the most
  likely process for the formation and evolution of the system.

\begin{acknowledgements}
This work made use of the Fourth Interferometric Catalogue, SIMBAD database and  CHORIZOS code of photometric and spectrophotometric data  analysis (http: //www.stsci.edu/ jmaiz/software/ chorizos/chorizos.html). The authors thanks Miss. Kawther Al-Waqfi for her help in some calculations and Mrs. Donna Keeley for the language editing.
\end{acknowledgements}


\end{document}